# Persistent type-II multiferroicity in nanostructured MnWO$_4$ ceramics

Pascaline Patureau[a], Rémi Dessapt[a], Philippe Deniard[a], U-Chan Chung[b,c], Dominique Michau[b,c], Michaël Josse[b,c], Christophe Payen[*,a], and Mario Maglione[*,b,c]

[a]Institut des Matériaux Jean Rouxel (IMN), UMR6502, Université de Nantes, CNRS, BP 32229, 44322 Nantes cedex 3, France

[b]CNRS, ICMCB, UPR 9048, F-33600 Pessac, France

[c]Univ. Bordeaux, ICMCB, UPR 9048, F-33600 Pessac, France

**ABSTRACT:** We show that the type-II multiferroic properties of bulk MnWO$_4$ are kept in nanostructured ceramics of ≈ 50 nm grain size prepared via spark plasma sintering. This means that ferroelectric polarization is robust against downsizing, which is at variance with standard ferroelectrics like BaTiO$_3$. We ascribe this stability to the spin-driven nature of the ferroelectric correlations in type-II multiferroics while it is resulting from lattice distortion in other cases. This may open the way for persistent type-II multiferroicity with no need for external stabilization like substrate-generated strain.

In archetypal ferroelectrics, the polarization and related properties result from structural lattice distortions[1]. For example, in BaTiO$_3$, it is the cationic off-centering versus the oxygen octahedra which is at the origin of the exceptional dielectric properties. Keeping these properties when downsizing such compounds is thus a challenge since the lattice elastic energy is disturbed by the surface energy even at relatively large grain size of several hundreds of nm[2,3,4]. In addition, depending on the processing conditions, lattice dipolar defects like –OH radicals are known to strongly interact with the lattice polarization. These issues may explain the large discrepancies in the literature about the nanosize effect on the polarization and transition temperature of standard three-dimensional (3D) ferroelectrics.[5] On the other hand, when the properties originate from electronic excitations like in magnetic, metallic and semi-conductor materials, the properties of 3D particles are disturbed at sizes below ≈ 10 nm. For example, high saturation moments and superparamagnetism were observed in maghemite (γ-Fe$_2$O$_3$) when particle sizes were less than 10 nm.[6]

Hoping to decrease the critical size for ferroelectricity, we thus want to probe the size effect in ferroelectrics whose polarization is related to electronic interactions. Magnetoelectric multiferroic materials in which the magnetic and ferroelectric orders coexist may be a good choice for this investigation. In this respect, the well-known type-I multiferroic BiFeO$_3$, which has lattice-related polarization resulting from Bi$^{3+}$ off-centering, was recently studied in the form of nanostructured ceramics prepared by Spark Plasma Sintering (SPS).[7] It was observed that both the Néel temperature and the ferroelectric Curie point shift to lower temperatures (from 340 °C to 265°C and from 697 °C to 655 °C°, respectively), and that piezoelectricity and ferroelectricity are still active at 40 nm scale. In the present work, we selected MnWO$_4$ because it is a well-known example of a type-II multiferroic in which ferroelectric transition and polarization result from magnetic ordering.[8,9] While previous studies have demonstrated the robustness of the ferroelectric transition against cationic substitution in MnWO$_4$[10,11], we have no knowledge of how a type-II multiferroic transition is affected by size effects. MnWO$_4$ has a simple monoclinic crystal structure which contains magnetic Mn$^{2+}$ ions and nonmagnetic W$^{6+}$ ions.[12] Three magnetic structures have been observed at low temperatures below $T_N$ = 13.5 K, $T_2$ = 12.5 K and $T_1$ ≈ 8 K.[12] Two of these magnetic structures are nonpolar states.[8,9] In the multiferroic state between $T_1$ and $T_2$, the so-called AF2 magnetic structure has an incommensurate magnetic unit cell of 2.8 nm$^3$. It is therefore plausible that multiferroicity survive finite-size effects in MnWO$_4$ particles of a few tens of nm. In this work, nanostructured MnWO$_4$ ceramics were prepared by SPS from nano-sized powders and characterized to check for type-II multiferroicity.

Nanopellets (P) and nanorods (R) of MnWO$_4$ were first synthesized in a reproducible manner under mild hydrothermal conditions according to the reported procedure of Chen *et al.*.[13] Samples were then briefly heated at 400 °C in air to eliminate water. Thermogravimetric curves of the as-obtained dried nanopowders showed only very small weight losses (≈ 0.5 %) at around 100 °C due to removal of adsorbed water (Supporting Information Figure S1). The purity and crystallinity of both types of nanoparticles were confirmed by X-ray diffraction (XRD) at room temperature (Supporting Information). Rietveld refine-ments were performed using the published $P2/c$ structural model of MnWO$_4$ (Supporting Information Figure S2 and Table S1)[12]. From this analysis, the computed crystallite sizes is 60 to 80 nm and 20 to 40 nm for MnWO$_4$-P and –R, respectively. The scanning electron microscopy (SEM) images of MnWO$_4$-P showed nanopellets with a mean diameter around 70 nm, while those of MnWO$_4$-R revealed uniform nanorods, with a mean diameter around 25 nm and a mean length of about 50 nm (Figure 1). In agreement with XRD results, this strongly suggests that each nanoparticle is a single crystallite.

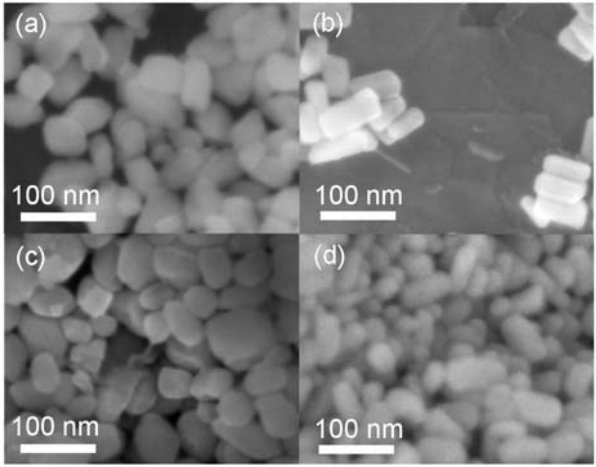

**Figure 1.** SEM images of (a) nanopellets of MnWO$_4$-P, (b) nanorods of MnWO$_4$-R, and of fracture surfaces in nanostructured SPS ceramics prepared at 520°C and 450 MPa using (c) MnWO4-P or (d) MnWO4-R nanopowders.

For SPS, the powders were loaded in a cylindrical tungsten carbide die with an inner diameter of 6 mm, internally coated with graphite paper and heated using an SPS equipment (Syntex Inc., Dr. Sinter SPS-515S). The heating was achieved owing to a pulsed direct electric current to cross the die. The use of an argon atmosphere instead of a vacuum one had no particular detectable effect in the process, as the heating and cooling rate, explored from 12.5°C/min to 50°C/min. A constant uniaxial pressure of 450 MPa was applied during the whole sintering process. Dense pellets were obtained by keeping the die at a constant temperature, in the range of 360°C to 520°C, for 10 min. For temperatures higher than 520°c, the shape and size of initial nanoparticles were found to be lost in the ceramics. The pellets were then gently abraded on SiC paper to remove any carbon surface contamination. Figure 1 shows SEM pictures for representative examples of nanostructured ceramics prepared at 520°C, revealing that both the shapes and the sizes of the grains are quite comparable to those of the initial nanopowders. Sintering bridges are unambiguously seen for both types of nanostructured ceramics. Because SPS ceramics have significant crystallographic texture, their XRD patterns were analyzed using the Le Bail method instead of Rietveld's one (Figure 2). The cell parameters and the crystallite sizes were found to be similar to those determined for MnWO$_4$-P and –R nanopowders (Supporting Information Table 1). The morphology, size and crystal structure of the initial nanoparticles are thus kept in dense nanoceramics thanks to SPS.

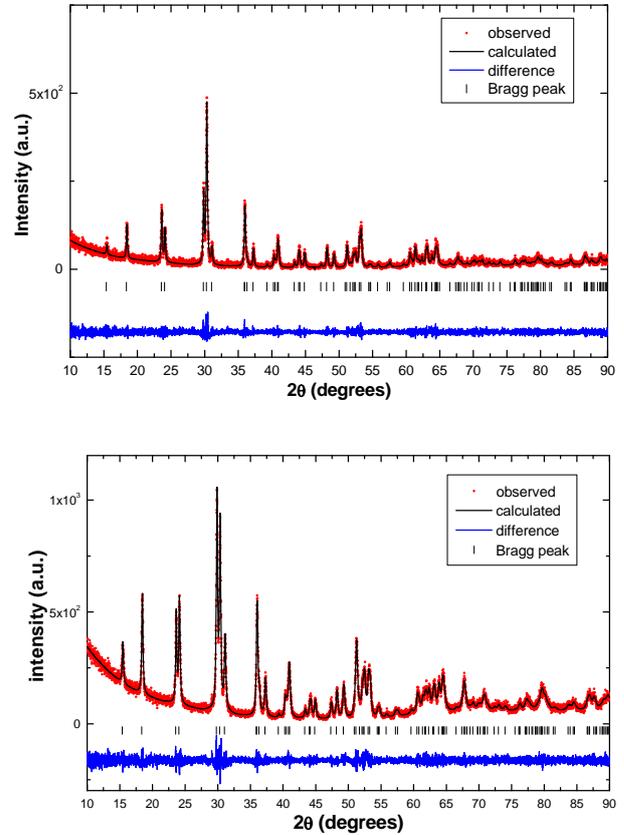

**Figure 2.** Room-temperature XRD patterns (taken with CuK-L$_3$ radiation) and Le Bail refinements for nanostructured MnWO4 ceramics prepared via SPS at 520°C and 450 MPa using MnWO4-P (top) or MnWO4-R (bottom) nanopowders.

Magnetic susceptibilities (DC or AC $\chi(T)$) and capacitance ($C(T)$) were then measured (Supporting Information) for both types of nanostructured dense pellets. The relative density of these pellets was ≈ 80%. We also measured dielectric loss factors but the signals were so low that we cannot separate them from the background. Dielectric properties were investigated in the range of 1 kHz-400 kHz. For the sake of comparison, DC-$\chi(T)$ and $C(T)$ data were also collected for a sintered sample prepared by heating a micrometric MnWO4 powder at high-temperature (HT; 1100 °C) in air and for a pellet sintered via SPS using the same micrometric powder. For these microstructured ceramics, crystallite sizes were found to be higher than the maximum measurable value ≈ 500 nm using our laboratory diffractometer. For all of the samples, Curie-Weiss fits to high-temperature DC-$\chi(T)$ data ($T > 100$ K) yielded effective magnetic moments par Mn atom, $\mu_{eff}$ ≈ 5.8 $\mu_B$, and Weiss temperatures, $\Theta$ ≈ -70 K, consistent with published values for micrometric powders or for single crystals.[8,11] Figure 3 shows both the DC-$\chi(T)$ and $C(T)$ curves at low temperatures. For the micrometric HT ceramic, DC-$\chi(T)$ is that of bulk MnWO$_4$ showing all three expected magnetic transitions at $T_1$, $T_2$ and $T_N$ (Figure 3a).[8,9] As usually done for a bulk three-dimensional antiferromagnet, transition temperatures are identified as the temperatures of peaklike anomalies in the d($\chi \cdot T$)/d$T$ versus $T$ curve.[14] Note that the actual Néel temperature $T_N$ = 13.45(5) K is a little bit smaller than the temperature at which a maximum occurs in DC-$\chi(T)$, $T(\chi_{max})$ = 13.70(5) K, and that the multiferroic transition at $T_2$ = 12.50(5) K is ≈ 1.2 K below the maximum in DC-$\chi(T)$. The frequency-independent peak in $C(T)$ at ≈ 12.5 K indicates that a ferroelectric transition occurs simultaneously at $T_2$.[11] The weak dielectric anomaly at $T_1$ is usually hardly detectable in ceramic samples.[8,9,11] As can be seen in Figure 3b, the microstructured SPS pellet has essentially the same behavior as that of the conventional HT ceramic. In contrast, the maximum in DC-$\chi(T)$ in all nanostructured samples slightly shifts to lower temperatures, $T(\chi_{max})$ = 13.3-13.4 K (Figure 3c-d). For each nanostructured sample, one does not observe any difference between the DC-$\chi(T)$ data measured under zero-field-cooled and that measured under field-cooled conditions (not shown), suggesting that the maximum in DC-$\chi(T)$ is not due to any spin-glass or cluster-glass or superparamagnetic behavior. Furthermore, the in-phase AC magnetic susceptibility, $\chi'(T)$, which is observed in the transition region, is frequency independent in the range of 0.10 – 1000 Hz (Supporting Information Figure S3). The maximum in $\chi'(T)$ occurs at the same temperature as in DC-$\chi(T)$. As can be seen in Figure S3, out-of-phase components of AC susceptibility, $\chi''(T)$, are constant and remain zero even below the maxima in $\chi'(T)$, indicating the absence of a dissipative process in the nanostructured samples.[15] These observations definitively indicate that the susceptibility maximum at ≈ 13.3-13.4 K is associated with a paramagnetic-to-antiferromagnetic transition lying somewhat below this maximum[14,15], as in bulk MnWO$_4$. In each nanostructured sample, a peak was observed in $C(T)$ at 11.7-11.8 K which is ≈ 1.6 K smaller than $T(\chi_{max})$ and a little bit smaller than bulk $T_2$. Because these peaks are frequency-independent in the investigated frequency range they clearly sign the persistence of ferroelectricity in both types of nanostructured SPS ceramics. Further evidence of multiferroicity in nanostructured ceramics was provided by the magnetic-field dependence of capacitance $C(T)$ (Figures 4 and S4). As in bulk MnWO$_4$,[9,16] the dielectric peak at $T_2$ is broadened and shifted toward lower temperature with increasing field $H$. The only visible qualitative difference in behavior between microstructured and nanostructured SPS samples is that the change of slope in the DC-$\chi(T)$ curve at $T_1$ is either suppressed or modified in the nanostructured pellets, depending on the grain morphology (Figure 3c-d). In this respect it is worth noting that the polar-to-nonpolar transition in bulk MnWO$_4$ at $T_1$ is associated both with a locking of the magnetic modulation with the lattice and with small discontinuous changes of lattice parameters.[12,17]

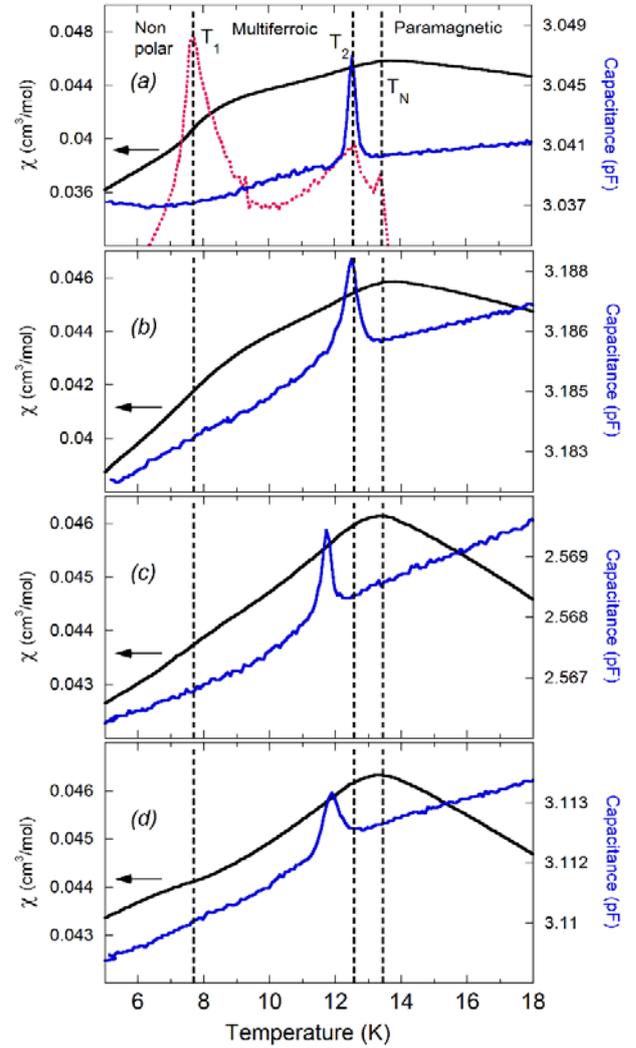

**Figure 3**. Temperature dependences of the DC magnetic susceptibility (solid black lines; left scales), $\chi$, and of the dielectric capacitance (solid blue lines; right scales), $C$, of microstructured MnWO$_4$ ceramics sintered (a) at high-temperature in air, (b) by SPS using a micrometric powder, and of nanostructured MnWO$_4$ ceramics prepared by SPS using (c) MnWO$_4$-P or (d) MnWO$_4$-R nanopowders. Vertical dashed lines indicate the transition temperatures $T_1$, $T_2$, and $T_N$ for the ceramic sintered at high-temperature in air, as determined from the d($\chi T$)/d$T$ versus $T$ curve shown by the dashed red line in panel (a) (d($\chi T$)/d$T$ is plotted in arbitrary unit not shown on the graph). Magnetic susceptibility data were acquired under a magnetic field of 1000 Oe. Capacitance data were collected at 385 kHz.

To summarize, nanostructured MnWO$_4$ ceramics have been successfully prepared via SPS for the first time, starting from nanopellet- and nanorod-like powders. A systematic study of SPS sintering parameters has allowed us to maintain grain and crystallite sizes for two types of grain morphology. Overall our data indicate that a prominent example of type-II multiferroic keeps showing the characteristic signatures of a multiferroic transition down to average grain sizes ≈ 50-70 nm for two grain morphologies. The robustness of type-II multiferroicity against downsizing in MnWO$_4$ is likely to be due to the spin-driven nature of the ferroelectric polarization as well as to the W-O chemical bonding that reduce effects caused by oxygen off-stoichiometry. In bulk MnWO$_4$, all oxygen atoms are actually strongly bonded to W atoms within WO$_6$ octahedra.[18] Others type-II multiferroics with polyanionic crystal structures like, e.g. silicates or germanates[19], may show the same remarkable behavior as the one reported here for MnWO$_4$. Our observations may thus pave the way for further investigations targeting the use of nanostructured type-II multiferroics in spintronic devices or sensors.

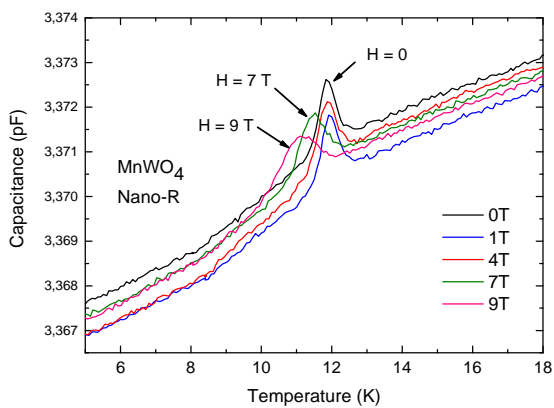

**Figure 4**. Temperature dependences of dielectric capacitance at selected magnetic fields (*H*) for a nanostructured ceramic prepared by SPS using MnWO$_4$-R nanopowders. Capacitance data were collected at 385 kHz.

**Corresponding Authors**
\* C. Payen. E-mail: christophe.payen@cnrs-imn.fr
\* M. Maglione. E-mail: Mario.Maglione@icmcb.cnrs.fr

**ACKNOWLEDGMENT**
We thank Nicolas Stephant and Rodolphe Decourt for their expert technical assistances.

**Supplementary material**

**Characterization methods**

<u>Laboratory powder X-ray diffraction.</u> X-ray diffraction (XRD) patterns were collected at room temperature on a Bruker D8 Advance instrument using monochromatic Cu$_{K-L3}$ ($\lambda$ = 1.540598 Å) X-rays and a LynxEye detector. Le Bail and Rietveld analyses of the XRD data were performed using JANA 2006 [Petricek, V.; Dusek, M.; Palatinus, L. Crystallographic Computing System JANA2006: General Features. Z. Für Krist. 2014, 229 (5), 345–352.] and the Cheary-Coelho fundamental approach for XRD profile parameters [Cheary, R. W.; Coelho, J. Appl. Crystallogr. 1998, 31 (6), 851–861 ; J. Appl. Crystallogr. 1998, 31 (6), 862–868.].

<u>Magnetic susceptibility.</u> A Quantum Design MPMS-XL7 equipped with an Evercool dewar was used to collect temperature-dependent DC and AC magnetization data. Zero field cooled (ZFC) and field cooled (FC) DC magnetization measurements were taken from 2 to 300 K in an applied field of $\mu_0 H$ = 0.1 T. Data were corrected for the diamagnetism of the sample holder as well as for core diamagnetism using Pascal's constants [Bain, G. A.; Berry, J. Chem. Educ. 2008, 85 (4), 532.].

<u>Dielectric permittivity.</u> Dielectric measurements were performed on dense pellets ($\approx$ 6 mm diameter, $\approx$ 1 mm thick) using an HP4194a impedance bridge. Samples were loaded into a Quantum Design Physical Properties Measurement System (PPMS). Measurements were taken in the frequency (f) range of 1 kHz–400 kHz and in the temperature range of 5 - 20 K.

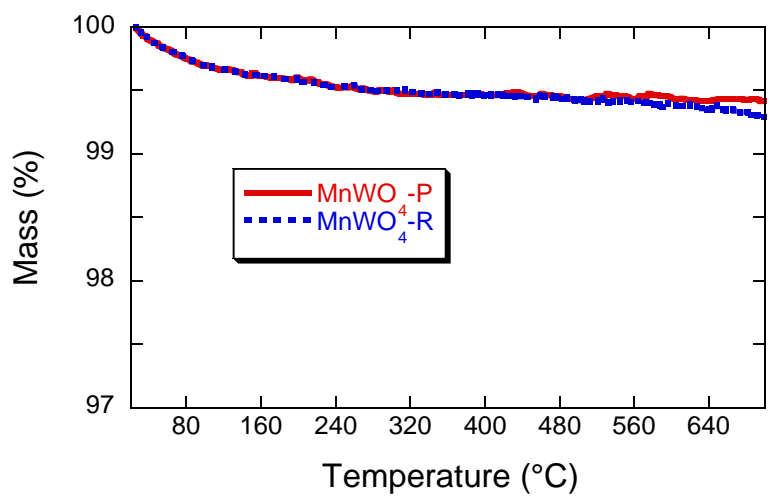

**Figure S1.** Thermogravimetric curves for MnWO$_4$ nanopowders in flowing air. The small weight loss ($\approx$ 0.5 %) at around 100° C is due to evaporation of adsorbed water.

**Table S1.** Cell parameters of nanopowders and of nanostructured ceramics of MnWO$_4$ as determined from Rietveld or Le Bail refinements of X-ray diffraction patterns recorded at room temperature. Refinements to the XRD data were made using the published *P*2/*c* structural model of MnWO$_4$ [Lautenschläger, G.; Weitzel, H.; Vogt, T.; Hock, R.; Böhm, A.; Bonnet, M.; Fuess, H. Phys. Rev. B: Condens. Matter Mater. Phys. 1993, 48, 6087−6098.].

|  | *a* (Å) | *b* (Å) | *c* (Å) | *β* (°) |
|---|---|---|---|---|
| MnWO$_4$-P nanopellets | 4.827(1) | 5.761(1) | 5.001(1) | 91.16(2) |
| MnWO$_4$-R nanorods | 4.826(2) | 5.758(4) | 4.996(1) | 91.19(1) |
| Nanostructured ceramic-P | 4.830(1) | 5.760(1) | 5.000(1) | 91.18(1) |
| Nanostructured ceramic-R | 4.829(1) | 5.758(2) | 5.000(1) | 91.21(3) |

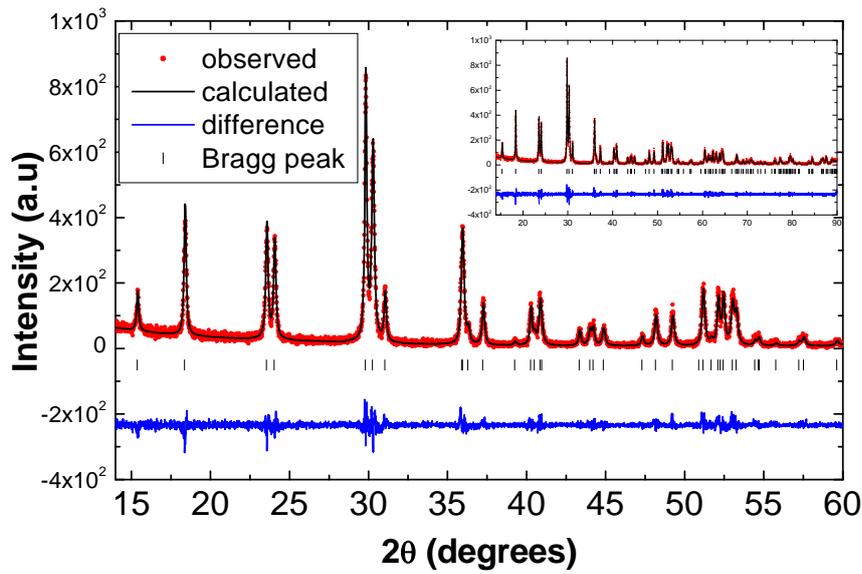

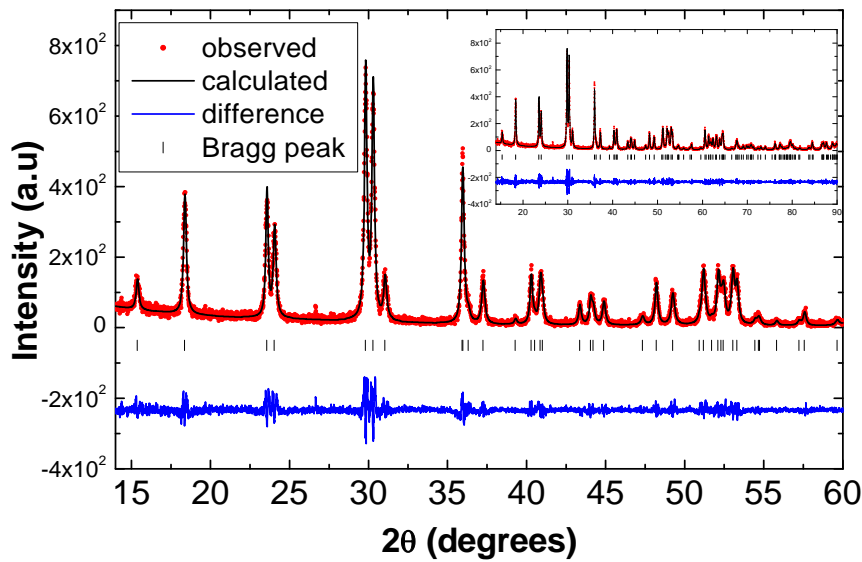

**Figure S2.** Final Rietveld refinement plots of the XRD data (λ = 1.540598 Å) for nanopellets MnWO$_4$-P (upper graph) and nanorods MnWO$_4$-R (lower graph) powders. The Rietveld refinements were obtained using the published *P*2/*c* structural model of MnWO$_4$ [Lautenschläger, G.; Weitzel, H.; Vogt, T.; Hock, R.; Böhm, A.; Bonnet, M.; Fuess, H. Phys. Rev. B: Condens. Matter Mater. Phys. 1993, 48, 6087−6098].

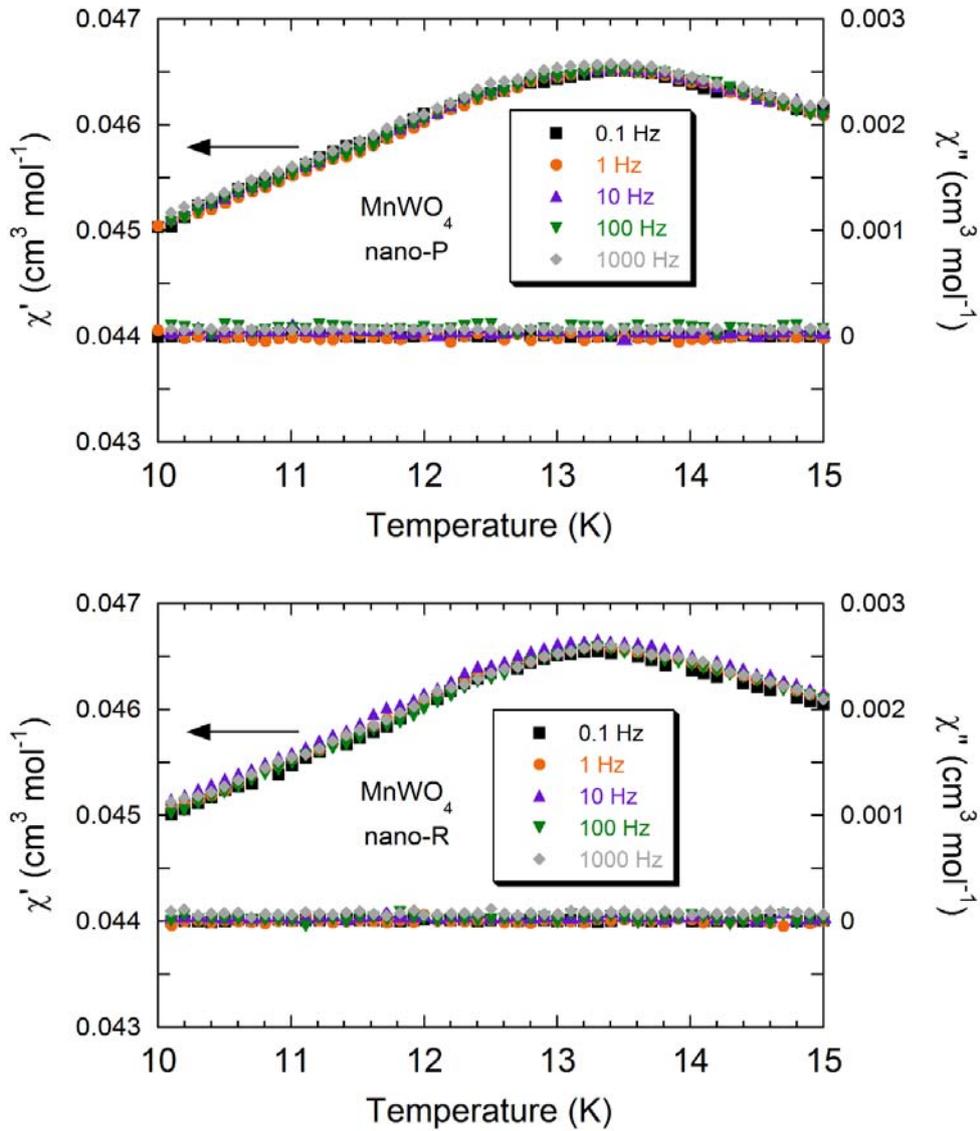

**Figure S3.** Temperature dependence of the in-phase, $\chi'$ (left scale), and out-of-phase, $\chi''$ (right scale), components of the AC susceptibility for nanostructured $MnWO_4$ ceramics prepared by SPS at 520 °C and 450 MPa using nanopellets $MnWO_4$-P (upper graph) or nanorods $MnWO_4$-R (lower graph). Data were taken at zero DC magnetic field for driving frequencies in the range of 0.1 Hz – 1000 Hz. Amplitude of the driving field was $H_{AC}$ = 3 Oe.

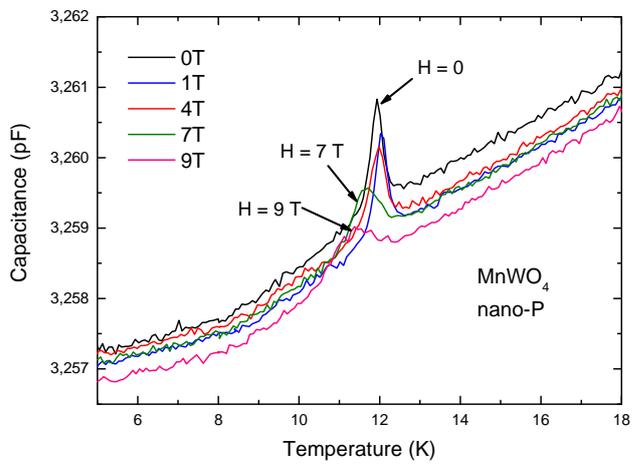

**Figure S4.** Temperature profiles of the dielectric capacitance at selected applied magnetic fields for nanostructured $MnWO_4$ ceramics prepared by SPS at 520 °C and 450 MPa using nanopellets $MnWO_4$-P. Data were collected at 385 Hz.